\documentclass[dvips,12pt]{article}
\usepackage{graphicx}
\newcommand{\doublespacing}{\let\CS=\@currsize\renewcommand{\baselinesstrech}
{2.0}\tiny\CS}

\begin{document}

\newcommand{\bc}{\begin{center}}
\newcommand{\ec}{\end{center}}
\newcommand{\bfr}{\begin{flushright}}
\newcommand{\efr}{\end{flushright}}
\newcommand{\lt}{\left}
\newcommand{\rt}{\right}
\newcommand{\vs}{\vspace}
\newcommand{\hs}{\hspace}

\bc {\bf \huge A classification  } \ec

\bc { \bf \huge of classical billiard trajectories  } \ec


\vspace{1cm}

\begin{center}
{\large \bf Bijan Bagchi $^{a,}${\footnote {e-mail :
bbagchi123@rediffmail.com, bbagchi123@gmail.com}} and
\bf Atreyee Sinha $^{b,}${\footnote {e-mail : atreyee.sinha@gmail.com}}} \\
\end{center}

\bc { $^a$ \it Department of Applied Mathematics, University of
Calcutta, \\ 92 Acharya Prafulla Chandra Road, Kolkata - 700 009,
INDIA} \ec

\bc { $^b$ \it St. Xavier's College, 30 Park Street, Kolkata - 700
016, INDIA} \ec

\vs{2cm}

\begin{abstract}

We examine the possible trajectories of a classical particle,
trapped in a two-dimensional infinite rectangular well, using the
Hamilton-Jacobi equation. We observe that three types of
trajectories are possible: periodic orbits, open orbits and some
special trajectories when the particle gets pocketed.

\end{abstract}

\vspace{3cm}

\pagebreak

\section{Introduction}

The so-called `Billiard systems', describing the motion of a
classical particle (a point ball) moving within a closed boundary
of different shapes, and bouncing perfectly from the walls, have
attracted the attention of various scientists for a long time
\cite{billiard}. Though the system appears simple, nevertheless it
is very rich and instructive, as the dynamics depends particularly
on the shape of the enclosure \cite{berry,kibble}. Consequently,
such systems have been studied both classically as well as in the
realm of quantum mechanics \cite{cq,asym}. It is assumed that
motion between collisions with the wall is in a straight line, and
at each bounce there is simple reflection with no dissipation,
i.e., the ball follows a path just like a light ray with a
boundary wall which is a perfect mirror \cite{berry}. Enclosures
of different shapes have been studied widely in the framework of
Hamilton-Jacobi theory and interesting results obtained. For
example, circular and elliptic enclosures with rigid boundaries
have been considered in \cite{berry} and the orbits traced out. In
particular, for such circular and elliptic enclosures a second
conserved quantity has been found other than the Hamiltonian,
leading to integrability and order. The trajectory for a circular
enclosure is found to be a succession of chords such that the
angular momentum of the particle about the centre remains constant
through successive bounce at the boundary. For the elliptical
enclosure the second conserved quantity is the product of the
angular momentum of the particle measured about the two foci of
the ellipse. In \cite{bb-cq} the periodic trajectories of a
particle trapped in an infinite square well have been explored,
using the Hamilton-Jacobi equation in 2-dimensions. Motivated by
such efforts, our aim in this work is to study the same for a
conventional billiard, modelled by an infinite rectangular well
potential.

\vspace{.5cm}

In section 2, we touch upon the Hamilton-Jacobi (H-J) equation
[3,7,8] and discuss the action-angle variables. In Section 3, we
apply the H-J equation to investigate the periodic classical
trajectories of a particle trapped inside a rectangular billiard
with infinite barriers. The different types of orbits, viz.,
periodic orbits, open trajectories, and those special trajectories
when the ball hits one of the corners and gets pocketed, are
discussed in detail in Section 4, with suitable illustrations.
Finally, Section 5 is kept for Conclusions and remarks.

\vspace{1cm}

\section{Hamilton-Jacobi Equation}

If $(q,p,t)$ is cannonically related to $(Q,P,t)$ under the
influence of a Hamiltonian $H(q,p,t)$ where $q=(q_1,q_2,...,q_n)$
and $p=(p_1,p_2,...,p_n)$ are generalized coordinates and
canonical momenta respectively, then, as is well known, the
Hamilton Jacobi equation is

\begin{equation}\label{HJ}
    \displaystyle \frac{\partial S}{\partial t} + H\left( q,
    \frac{\partial S}{ \partial q} , t \right) = 0
\end{equation}

where S is the  \textit{Hamilton's principal function}

Writing ,
\begin{equation}\label{s}
    S = W(q_1,q_2, \cdots , q_n) - \alpha _1 t
\end{equation}
where the time-independent function $W(q)$ is  {\it Hamilton's
characteristic function} and the constant $ \alpha _1$ is the
energy $E$. This transforms the H-J equation to
\begin{equation}\label{he}
    H \left( q, \frac{\partial S}{\partial q} \right) = E
\end{equation}

For integrable systems there is a natural set of co-ordinates and
momenta which is particularly convenient and useful
\cite{kibble,goldstein}. These systems have $n$ distinct constants
of motion and we can transform to a new set of coordinates $w_i$
and momenta $J_i$ in such a way that \\
$*$ the Hamiltonian form of the equations of motion is preserved : $ K(J_j) \equiv H $, \\
$*$ the new momenta $J_i$ are all constants of motion, \\
$*$ the new coordinates $ w_i$ are all ignorable. \\
Writing the Hamilton's characteristic function as $ W = W( q, J)
$, the {\it angle} variables $w_i$ and the corresponding
canonically conjugate {\it action} variables $J_i$ are given by
\begin{equation}\label{angle}
    \displaystyle \dot{w}_j ~=~ \displaystyle \frac{\partial K(J)}{\partial J_j}
    ~ \equiv ~ \nu(J)
\end{equation}
\begin{equation}\label{action}
    \displaystyle \dot{J}_j ~=~ \displaystyle - \frac{\partial K}{\partial w_j}
    ~ = ~ 0
\end{equation}
For Hamilton's equations to be form-invariant it is necessary that
the change of variables $(q,p) \rightarrow (w,J)$ should preserve
areas, so that a natural choice for $J$ is given by
\begin{equation}\label{i}
    J = \displaystyle \oint pdq
\end{equation}
where the integration is carried over a complete period of
libration or rotation, as the case may be. The {\it angle
variable}, which is the generalized coordinate conjugate to $J$,
is defined by the transformation equation
\begin{equation}\label{w}
    w(J) = \displaystyle \frac{\partial W}{\partial J}
\end{equation}
It is evident from (\ref{angle}) that the angle variables evolve
at a uniform rate, given by
\begin{equation}\label{wnu}
    w_j = \nu _j t + \beta _j
\end{equation}
where $\nu _j $ is the {\it frequency} associated with the
periodic motion and $ \beta _j$ are constants. The use of
action-angle variables thus provides a powerful technique for
obtaining the frequency of periodic motion without finding a
complete solution to the motion of the system.

\vspace{1cm}

\section{Motion of a particle in a 2-dim. rectangular well}

With the above background, let us investigate the nature of the
trajectories when the particle is in an infinite rectangular well,
with centre at $(0,0)$ and vertices at $( \pm a \ , \ \pm b)$ :
\begin{equation}\label{v2d}
    V(x,y)= \left\{
    \begin{array}{lcl}
    0 \qquad \ \ \ \ \ \ \ |x| < a  \qquad {\rm{and}} \qquad |y| < b
    \\ \\
    \infty \qquad \ \ \ \ \ \  |x| > a  \qquad {\rm{and}} \qquad |y| > b \\
    \end{array}
    \right.
\end{equation}
A canonical transformation from $(q,p) \rightarrow (Q,P)$, i.e.,
to new variables $( \beta , \alpha )$ which are constants in time,
and, employing a type-two generating function $F_2 (q,P,t) \equiv
S(q, \alpha , t) $, with the assumption that $ H $ does not depend
on $t$ explicitly, (\ref{he}) reduces to the form
\begin{equation}\label{we}
    \displaystyle \frac{1}{2m} \left[ \left( \frac{\partial
    W}{\partial x} \right) ^2 + \left( \frac{\partial
    W}{\partial y} \right) ^2 \right] = E
\end{equation}
Writing $ W(q) = X(x) + Y(y) $ in (\ref{we}), the constants of
motion are obtained as
\begin{equation}\label{alpha-xy}
    \displaystyle \left( \frac{dX}{dx} \right) ^2  = \alpha _x ^2
    \qquad , \qquad
    \displaystyle \left( \frac{dY}{dy} \right) ^2  = \alpha _y ^2
\end{equation}
with $ \displaystyle E = \frac{1}{2m} \left( \alpha _x ^2 + \alpha
_y ^2 \right) $, yielding
\begin{equation}\label{pxy}
    p_x = \pm \alpha _x \qquad , \qquad p_y = \pm \alpha _y
\end{equation}
the signs showing the reversal in the direction of motion of the
particle each time it hits the barriers at $x = \pm a$ and
$ y = \pm b$. \\
The action variables can now be calculated easily from (\ref{i})
and turn out to be
\begin{equation}\label{Jx}
    J_x = 4 a \alpha _x
\end{equation}
\begin{equation}\label{Jy}
    J_y = 4b \alpha _y
\end{equation}
so that $E$ becomes
\begin{equation}\label{e}
    E = \displaystyle \frac{1}{ 32 m } \left( \frac{J_x ^2}{
    a^2} + \frac{J_y ^2}{b^2} \right)
\end{equation}
Thus the natural frequencies of the system are obtained from
(\ref{angle}) to be
\begin{equation}\label{wxy}
    \nu _x = \displaystyle \frac{\partial E}{ \partial J_x} =
    \frac{J_x }{ 16 m a^2} \qquad , \qquad
    \nu _y = \displaystyle \frac{\partial E}{ \partial J_y} =
    \frac{J_y }{ 16 m b^2}
\end{equation}
which, with the help of (\ref{Jx}) and (\ref{Jy}) read
equivalently
\begin{equation}\label{nuxy}
    \nu _x = \displaystyle  \frac{p_x}{4ma} \qquad , \qquad \nu _y
    = \frac{p_y}{4mb}
\end{equation}
Thus the natural frequencies are functions of the particle
velocity and the dimensions of the well. Consequently, three types
of trajectories are possible for the particle trapped in the
rectangular well, as discussed below.

\vspace{1cm}


\section{Possible Trajectories}

In this section we shall discuss in detail the three possible
trajectories of the trapped particle, viz., \\
1. Periodic Trajectories \\
2. Open Trajectories \\
3. Special Trajectories when the particle hits one of the corners
and gets pocketed.

\subsection{Periodic Trajectories :}

\vspace{.4cm}

One of our primary aims in this work is to study periodic or
closed trajectories. It is evident from equation (\ref{nuxy}), for
the particle to execute periodic motion, it must return to the
starting point with its initial momenta after a certain time. This
is possible only if
\begin{equation}\label{t}
    T = n_x T_x = n_y T_y
\end{equation}
where $T_x$ and $T_y$ represent the time in which the particle
reaches the starting point with its initial momenta in the $x$ and
$y$ directions respectively, $T$ is the time period of the orbit,
and $n_x \ , \ n_y$ are integers. Thus, if the particle starts
from the origin at an angle $ \theta $ to the $x$ direction, where
\begin{equation}\label{theta}
    \displaystyle \tan \ \theta = \frac{p_y}{p_x}
\end{equation}
then, for closed orbits $\tan \ \theta$ must be rational
{\footnote{We consider those cases where the linear momentum in
the $x$ and $y$ directions, viz., $p_x$ , $p_y$ are expressible in
rational form, so that the action variables $J_x$ , $J_y$ , as
well as the natural frequencies $ \nu _x $ , $\nu _y$ are also
rational.}}, with the time period given by (\ref{t}). With the
help of (\ref{nuxy}), eq. (\ref{theta}) may be rearranged to give
\begin{equation}\label{theta-p}
    \tan \ \theta = \displaystyle
    \frac{b \nu _y}{a \nu _x} = \frac{b T_x}{a T_y} =
    \frac{b n_y}{a n_x}
\end{equation}
We shall illustrate this with a couple of explicit examples below,
the particle starting from the origin in each case. The
corresponding closed orbits are plotted in Figures 1 and 2; the
starting trajectory is shown in blue, the intermediate ones in
red, and the closing one in black.

\vspace{.5cm}


\subsubsection{Some explicit examples for periodic trajectories}

\noindent {\bf Case 1 : $n_y = 1, n_x = 4$ , i.e.,  $
\displaystyle \frac{p_y}{p_x} = \frac{b}{4a} $ } \\

\noindent This case is illustrated in Figure 1. The trajectories
traced out by the particle are

\begin{equation}\label{2d2}
\begin{array}{lcl}
    x &=& \displaystyle \frac{p_x}{m}t \qquad
    \qquad \qquad \qquad \qquad \
    y = \displaystyle \frac{p_y}{m}t \qquad
    \qquad \qquad \qquad \qquad  \ \ \ t_{16 (r -1)} \leq t \leq t_{16 r -15}
    \\ \\
    x &=& \displaystyle - \frac{p_x}{m} \left( t - t_{16 r -15}
    \right) + a  \qquad \ \ \
    y = \displaystyle  \frac{p_y}{m} \left( t - t_{16 r -15}
    \right) + \frac{b}{4}  \qquad \ \ \ \ \ t_{16 r -15} \leq t \leq
    t_{16 r -14 }
    \\ \\
    x &=& \displaystyle - \frac{p_x}{m} \left( t- t_{16 r -14} \right)
    \qquad \qquad \  \
    y = \displaystyle \frac{p_y}{m} \left( t - t_{16 r -14}
    \right) + \frac{b}{2} \qquad \ \ \ \ \
    t_{16 r - 14} \leq t \leq t_{16 r -13}
    \\ \\
    x &=& \displaystyle \frac{p_x}{m} \left( t - t_{{16 r -13}} \right) -a
    \qquad \qquad
    y = \displaystyle  \frac{p_y}{m} \left( t - t_{16 r -13}
    \right) + \frac{3b}{4} \qquad \ \ \ \
    t_{16 r - 13} \leq t \leq t_{16 r - 12}
    \\ \\
    x &=& \displaystyle \frac{p_x}{m} \left( t- t_{16 r -12} \right)
    \qquad \qquad \ \ \ \ \
    y = \displaystyle  - \frac{p_y}{m} \left( t - t_{16 r -12}
    \right) + b  \qquad \ \ \ \
    t_{16 r - 12} \leq t \leq t_{16 r - 11}
    \\ \\
    x &=& \displaystyle - \frac{p_x}{m} \left( t- t_{16 r -11} \right)
    + a \qquad \ \ \ \
    y = \displaystyle - \frac{p_y}{m} \left( t - t_{16 r -11}
    \right) + \frac{3b}{4}  \qquad \ \
    t_{16 r - 11} \leq t \leq t_{16 r -10}
    \\ \\
    x &=& \displaystyle - \frac{p_x}{m} \left( t- t_{16 r -10} \right)
    \qquad \qquad \ \ \
    y = \displaystyle - \frac{p_y}{m} \left( t - t_{16 r -10}
    \right) + \frac{b}{2}  \qquad \ \ \ \
    t_{16 r - 10} \leq t \leq t_{16 r -9}
    \\ \\
    x &=& \displaystyle \frac{p_x}{m} \left( t - t_{16 r -9}
    \right) - a  \qquad \qquad \ \
    y = \displaystyle  - \frac{p_y}{m} \left( t - t_{16 r -9}
    \right) + \frac{b}{4}  \qquad \ \ \ \ t_{16 r -9} \leq t \leq
    t_{16 r -8 }
    \\ \\
    x &=& \displaystyle \frac{p_x}{m} (t - t_{16 r - 8}) \qquad
    \qquad \qquad  \ \
    y = \displaystyle - \frac{p_y}{m} (t - t_{16 r - 8}) \qquad
    \qquad \ \ \ t_{16 r - 8} \leq t \leq t_{16 r -7}
    \\ \\
    x &=& \displaystyle - \frac{p_x}{m} \left( t - t_{16 r -7}
    \right) + a  \qquad \ \ \ \
    y = \displaystyle - \frac{p_y}{m} \left( t - t_{16 r -7}
    \right) - \frac{b}{4}  \qquad \ \ \ \ t_{16 r -7} \leq t \leq
    t_{16 r -6 }
    \\ \\
    x &=& \displaystyle - \frac{p_x}{m} \left( t- t_{16 r -6} \right)
    \qquad \qquad \  \ \
    y = \displaystyle - \frac{p_y}{m} \left( t - t_{16 r -6}
    \right) - \frac{b}{2} \qquad \ \ \ \ \
    t_{16 r - 6} \leq t \leq t_{16 r -5}
    \\ \\
    x &=& \displaystyle \frac{p_x}{m} \left( t - t_{{16 r -5}} \right) -a
    \qquad \qquad \
    y = \displaystyle  - \frac{p_y}{m} \left( t - t_{16 r -5}
    \right) - \frac{3b}{4} \qquad \ \ \ \
    t_{16 r - 5} \leq t \leq t_{16 r -4}
    \\ \\
    x &=& \displaystyle \frac{p_x}{m} \left( t- t_{16 r -4} \right)
    \qquad \qquad \ \ \ \ \ \
    y = \displaystyle  \frac{p_y}{m} \left( t - t_{16 r -4}
    \right) - b  \qquad \ \ \ \
    t_{16 r - 4} \leq t \leq t_{16 r - 3}
    \\ \\
    x &=& \displaystyle - \frac{p_x}{m} \left( t- t_{16 r -3} \right)
    - a \qquad \ \ \ \ \
    y = \displaystyle \frac{p_y}{m} \left( t - t_{16 r -3}
    \right) -  \frac{b}{4}  \qquad \ \
    t_{16 r - 3} \leq t \leq t_{16 r -2}
    \\ \\
    x &=& \displaystyle - \frac{p_x}{m} \left( t- t_{16 r -2} \right)
    \qquad \qquad \ \ \ \
    y = \displaystyle  \frac{p_y}{m} \left( t - t_{16 r -2}
    \right) - \frac{b}{2}  \qquad \ \ \ \
    t_{16 r - 2} \leq t \leq t_{16 r -1}
    \\ \\
    x &=& \displaystyle \frac{p_x}{m} \left( t - t_{16 r -1}
    \right) - a  \qquad \qquad \ \
    y = \displaystyle   \frac{p_y}{m} \left( t - t_{16 r -1}
    \right) - \frac{b}{4}  \qquad \ \ \ \ t_{16 r -1} \leq t \leq
    t_{16 r  }
    \\ \\
\end{array}
\end{equation}
where $r = 1, 2, \cdots$ and $ t_n = \displaystyle (2n -1)
\frac{ma}{p_x} \ , \ n = 1 ,2 , \cdots $
 \\
Thus, in this case, the trajectories are periodic when the
velocities of the particle in the $x$ and $y$ directions are such
that the time taken by the particle to cover distance $a$ in the
$x$ direction, is the same as the time it takes to cover the
distance $b/4$ in the $y$ direction :
    $$ \displaystyle t_1 = \frac{ma}{p_x} = \frac{mb}{4p_y} $$
This gives the time period as $ T = 4 T_x = T_y $, where $ T_x = 4
t_1 $ and $ T_y = 16 t_1 $. This is a new result not discussed for
$a=b$ in ref. \cite{bb-cq}.

\vspace{1cm}


\noindent {\bf Case 2 : $n_y = 2, n_x = 3$ , i.e.,  $
\displaystyle \frac{p_y}{p_x} = \frac{2b}{3a} $ } \\

\noindent This case is illustrated in Figure 2. The particle can
be shown to trace out the following trajectories :

\begin{equation}\label{2d3}
\begin{array}{lcl}
    x &=& \displaystyle \frac{p_x}{m}t \qquad
    \qquad \qquad \qquad \qquad \
    y = \displaystyle \frac{p_y}{m}t \qquad
    \qquad \qquad \qquad \qquad  \ \ \ t_{12 (r -1)} \leq t \leq t_{12 r -11}
    \\ \\
    x &=& \displaystyle - \frac{p_x}{m} \left( t - t_{12 r -11}
    \right) + a  \qquad \ \ \
    y = \displaystyle \frac{p_y}{m} \left( t - t_{12 r -11}
    \right) + \frac{2b}{3}  \qquad \ \ \  t_{12 r -11} \leq t \leq
    t_{12 r -10 }
    \\ \\
    x &=& \displaystyle - \frac{p_x}{m} \left( t- t_{12 r -10} \right)
    + \frac{a}{2} \qquad \ \ \
    y = \displaystyle - \frac{p_y}{m} \left( t - t_{12 r -10}
    \right) + b \qquad \ \ \ \ \
    t_{12 r - 10} \leq t \leq t_{12 r -9}
    \\ \\
    x &=& \displaystyle \frac{p_x}{m} \left( t - t_{{12 r -9}} \right) -a
    \qquad \qquad \
    y = \displaystyle  - \frac{p_y}{m} \left( t - t_{12 r -9}
    \right)  \qquad \qquad
    t_{12 r - 9} \leq t \leq t_{12 r - 8}
    \\ \\
    x &=& \displaystyle \frac{p_x}{m} \left( t- t_{12 r -8} \right)
    + \frac{a}{2} \qquad \qquad  \
    y = \displaystyle   \frac{p_y}{m} \left( t - t_{12 r -8}
    \right) - b  \qquad \ \ \ \
    t_{12 r - 8} \leq t \leq t_{12 r - 7}
    \\ \\
    x &=& \displaystyle - \frac{p_x}{m} \left( t- t_{12 r -7} \right)
    + a \qquad \ \ \ \
    y = \displaystyle \frac{p_y}{m} \left( t - t_{12 r -7}
    \right) - \frac{2b}{3}  \qquad \ \
    t_{12 r - 7} \leq t \leq t_{12 r -6}
    \\ \\
    x &=& \displaystyle - \frac{p_x}{m} \left( t- t_{12 r -6} \right)
    \qquad \qquad \ \ \
    y = \displaystyle \frac{p_y}{m} \left( t - t_{12 r -6}
    \right)   \qquad \qquad \ \ \
    t_{12 r - 6} \leq t \leq t_{12 r -5}
    \\ \\
    x &=& \displaystyle \frac{p_x}{m} \left( t - t_{12 r -5}
    \right) - a  \qquad \qquad \
    y = \displaystyle   \frac{p_y}{m} \left( t - t_{12 r -5}
    \right) + \frac{2b}{3}  \qquad \ \ \ \ t_{12 r -5} \leq t \leq
    t_{12 r -4 }
    \\ \\
    x &=& \displaystyle \frac{p_x}{m} (t - t_{12 r - 4}) - \frac{a}{2}
    \qquad
    \ \ \ \
    y = \displaystyle - \frac{p_y}{m} (t - t_{12 r - 4}) - b \qquad
    \qquad \ \ \ t_{12 r - 4} \leq t \leq t_{12 r -3}
    \\ \\
    x &=& \displaystyle - \frac{p_x}{m} \left( t - t_{12 r -3}
    \right) + a  \qquad \ \ \ \ \
    y = \displaystyle - \frac{p_y}{m} \left( t - t_{12 r -3}
    \right)  \qquad \ \ \ \ t_{12 r -3} \leq t \leq
    t_{12 r -2 }
    \\ \\
    x &=& \displaystyle - \frac{p_x}{m} \left( t- t_{12 r -2}
    \right)- \frac{a}{2}
    \qquad \ \  \ \
    y = \displaystyle \frac{p_y}{m} \left( t - t_{12 r -2}
    \right) - b \qquad \ \ \ \ \
    t_{12 r - 2} \leq t \leq t_{12 r -1}
    \\ \\
    x &=& \displaystyle \frac{p_x}{m} \left( t - t_{{12 r -1}} \right) -a
    \qquad \qquad \
    y = \displaystyle   \frac{p_y}{m} \left( t - t_{12 r -1}
    \right) - \frac{2b}{3} \qquad \ \ \ \
    t_{12 r - 1} \leq t \leq t_{12 r }
    \\ \\
\end{array}
\end{equation}
where $r = 1, 2, \cdots$ and $ t_n = \displaystyle (2n -1)
\frac{ma}{p_x} \ , \ n = 1 ,2 , \cdots $
 \\ \\
It is easy to observe from Figure 2 that in this case $t_2 = 3t_1
/2  , \ t_3 = 3 t_1  , \ t_4 = 9t_1 / 2  , \ t_5 = 5 t_1 $, etc.
Thus $ T_x = t_5 - t_1 = \displaystyle 4 t_1 $ and $ T_y = t_8 -
t_2 = \displaystyle 6 t_1 $, where $ t_1 = \displaystyle
\frac{ma}{p_x} = \frac{2mb}{3 p_y}$, giving $ T = 12 t_1 = 3 T_x =
2 T_y $. Note that we get back the result of \cite{bb-cq} for
$a=b$.


\begin{figure}[hp]
\begin{center}
  \includegraphics[width=18cm]{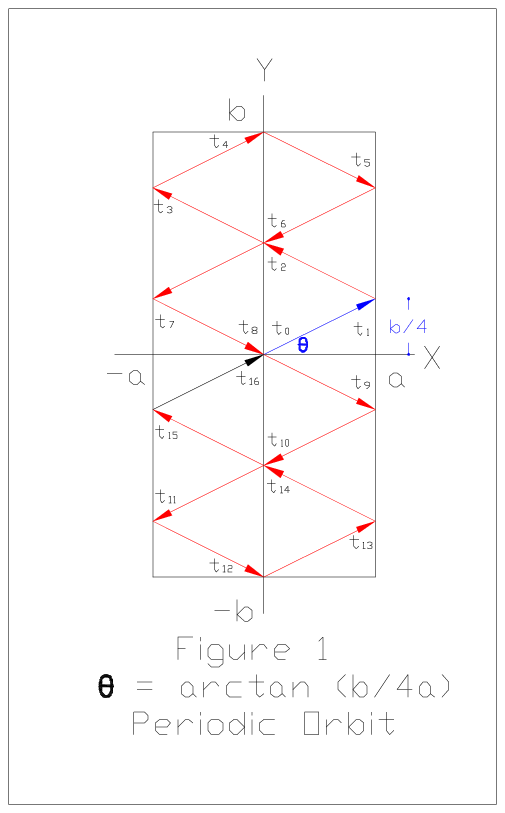}
  \caption{\small This shows a periodic trajectory for
  $ \displaystyle n_y = 1 , \ n_x = 4$, so that
  $ \tan \theta = \displaystyle \frac{p_y}{p_x} = \frac{b}{4a}$}\label{}
\end{center}
\end{figure}


\vspace{.5cm}


{\begin{figure}[hp]
\begin{center}
\includegraphics[width=18cm]{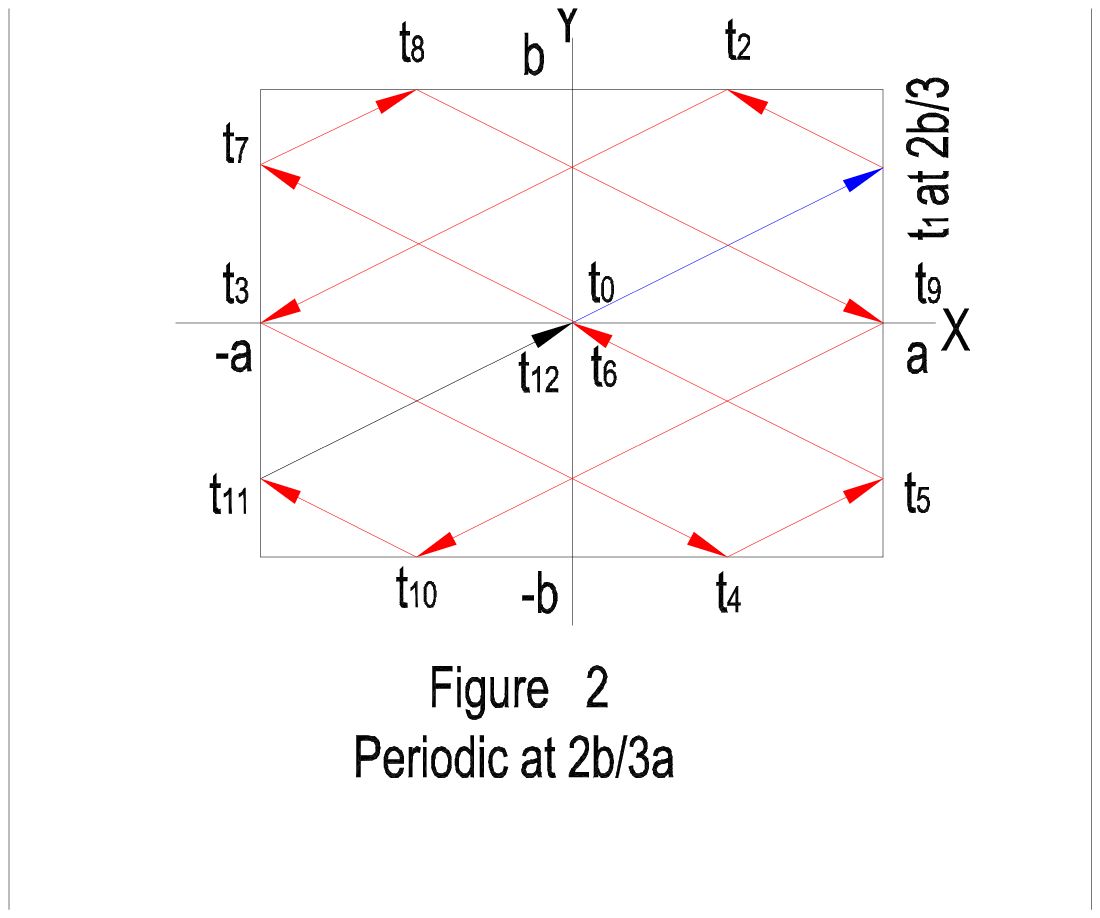}
\label*{}\caption{\small This shows a periodic trajectory for
$\displaystyle n_y = 2, n_x = 3$ , so that $ \tan \theta =
\displaystyle \frac{p_y}{p_x} = \frac{2b}{3a}$ }
\end{center}
\end{figure}}

\pagebreak

\subsection{Open Trajectories :}

\vspace{.5cm}

If the initial angle $ \theta $ is such that $ \displaystyle
\frac{p_y}{p_x} $ is irrational, then the orbit is an open one.
This is due to the fact that the time periods in the $x$ and $y$
directions are such that one cannot find integral values of $n_x$
, $n_y$ satisfying equation (\ref{t}). We have traced out such a
trajectory in Figure 3, for $ \theta = 30 ^0 $, starting with the
blue line. Even after multiple reflections from the perfectly
elastic walls of the rectangular well (shown by the red paths) the
orbit does not close as is evident from the black line with the
arrowhead.

\vspace{.5cm}


\subsection{Special Trajectories when the billiard ball hits a
corner:}

\vspace{.5cm}

We now address the interesting case of special non-periodic
trajectories when the particle hits one of the corners of the
billiard table and gets pocketed. For the ball to hit either the
right or left wall, the distance travelled in the $x$ direction is
$ (4n \pm 1)a $, where $n$ is an integer. Similarly, to hit the
top or bottom wall the distance travelled in the $y$ direction is
$ (4m \pm 1)b $, where $m$ is also an integer. If the ball hits a
corner then these two conditions must be satisfied simultaneously.
Thus, if the particle hits a corner in time $t$, then
\begin{equation}\label{corner}
    \displaystyle \frac{v_y t}{v_x t} = \displaystyle
    \frac{\left(4m \pm 1 \right) b}{\left(4n \pm 1 \right) a}
\end{equation}
and the condition for the billiard ball to get pocketed reduces to
\begin{equation}\label{pocket}
    \displaystyle \frac{p_y}{p_x} = \displaystyle
    \frac{\left(4m \pm 1 \right) b}{\left(4n \pm 1 \right) a}
\end{equation}
If the numerator has $+ve$ ($-ve)$) sign in (\ref{pocket}), then
the ball hits one of the two corners where $y$ is positive
(negative). Similarly, if the denominator has $+ve$ ($-ve)$) sign
in (\ref{pocket}), then the ball hits one of the two corners where
$x$ is positive (negative). From equations (\ref{theta}),
(\ref{theta-p}) and (\ref{pocket}), it is obvious that this occurs
for odd integral values of both $n_x$ and $n_y$. Based on this we
summarize in Table 1 the corner in which the ball will get
pocketed. It may be mentioned that we assign the following numbers
to the respective corners : $(a,b)$ as corner 1, $(-a,b)$ as
corner 2, $(-a,-b)$ as corner 3, and $(a,-b)$ as corner 4. We
shall trace out the actual trajectories in Fig 4. It is observed
that the predicted corners are in fact the actual ones.


\pagebreak

\begin{center}
{\bf Table 1 :}

\vspace{.25cm}

\begin{tabular}{|c|c|c|}
  \hline
   & & \\
   $\displaystyle n_y / n_x $ & sign (numerator, denominator) & corner \\
   & & \\ \hline
   $ 1/3 $ & $(-,+)$ & 2 \\
   \hline
   $ 1/5 $ & $(+,+)$ & 1 \\
   \hline
   $ 1/7 $ & $(-,+)$ & 2 \\
   \hline
   $ 1/9 $ & $(+,+)$ & 1 \\
   \hline
   $ 3/7 $ & $(-,-)$ & 3 \\
   \hline
   $ 5/7 $ & $(-,+)$ & 2 \\
   \hline
   $ 3/5 $ & $(+,-)$ & 4 \\
   \hline
   $ 5/9 $ & $(+,+)$ & 1 \\
   \hline
   $ 7/9 $ & $(+,-)$ & 4 \\
   \hline
\end{tabular}
\end{center}

\vspace{.7cm}


\begin{figure}[hp]
\begin{center}
  \includegraphics[width=21cm]{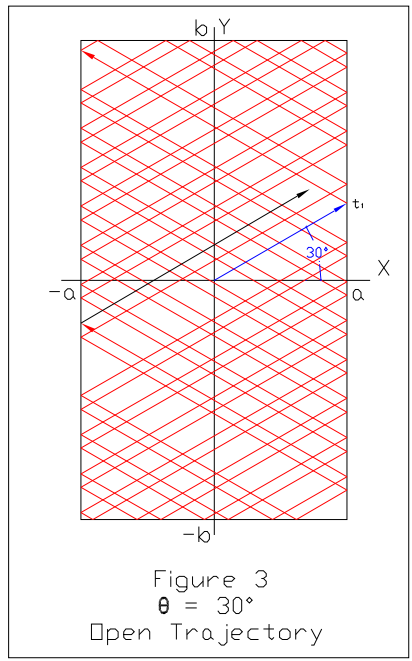}
  \caption{\small This shows an open trajectory for
  $\theta = 30 ^o$}\label*{}
\end{center}
\end{figure}

\pagebreak

\begin{figure}[hp]
\begin{center}
  \includegraphics[width=21cm]{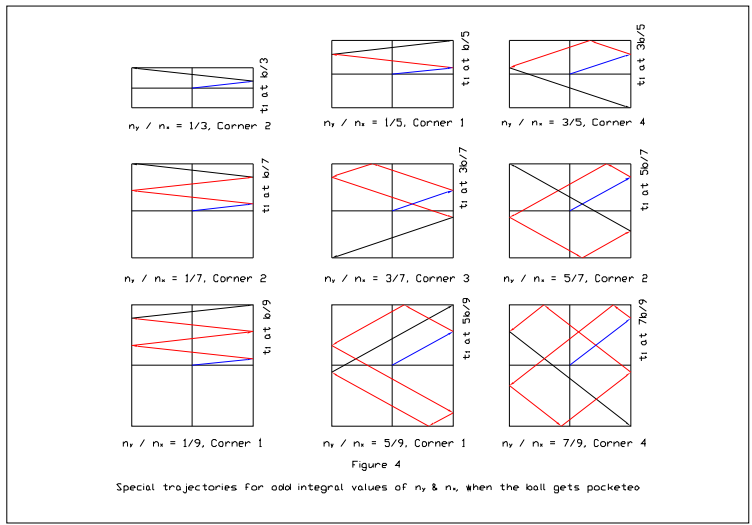}
  \caption{\small This shows some of the special trajectories for odd integral
values of both $n_y$ and $n_x$ , when the billiard ball hits one
of the corners and gets pocketed. The corners predicted in Table 1
agree with those actually traced out in this figure} \label*{}
\end{center}
\end{figure}

\pagebreak

\section{Conclusions }

To conclude, we have studied the motion of a classical point
particle, trapped in an infinite rectangular well with perfectly
elastic boundaries, using the action-angle variables in the
Hamilton-Jacobi formalism. In particular, we have determined the
natural frequencies of the system in the $x$ and $y$ directions.
These frequencies given by $\nu _x$, $ \nu _y$ in eq. (\ref{wxy}),
are found to be functions of the velocity of the trapped particle,
and the dimensions of the rectangular well, viz., $a, \ b$
respectively. It may be worth mentioning here that since the
potential $V = 0$ inside the enclosure, the magnitudes of the
particle momenta in the $x$ and $y$ directions ($p_x$ and $p_y$)
do not change inside the well. We have established a definite
relationship between the orbit traced out by the classical
particle and the initial angle (say $\theta$ with the $x$ axis) at
which the particle starts from rest from the origin, i.e., on the
ratio $ p_y / p_x $. When $p_x$ , $p_y$ are both rational,
depending on the values of $n_x$ and $n_y$, some orbits with
definite periodicity have been illustrated in Figures 1 and 2. In
these cases one can find integral values of $n_x \ , \ n_y$ for
which the relationship $n_x T_x = n_y T_y $ holds, and the time
period of the periodic motion is obtained as $ T = n_x T_x = n_y
T_y $.

On the other hand, if the initial angle $ \theta $ is such that $
\tan \ \theta = p_y / p_x $ is irrational, the orbit is an open
one. Such an open orbit has been sketched in Figure 3, for the
particular value of $ \theta = 30 ^0$.

Still more interesting are the cases when the billiard ball falls
into one of the pockets. We have shown that this occurs for odd
integral values of both $n_x$ and $n_y$. A few such trajectories
are plotted in Figure 4. In fact, our conjecture can even predict
accurately which corner the particle would hit. The excellent
agreement between Table 1 and Fig 4 gives credence to our
conjecture.

Additionally, we have observed that the qualitative picture does
not depend on the dimensions of the well, i.e. on the ratio $
a/b$. For the special case $a=b$, i.e., an infinite square well,
our results reduce to those of ref \cite{bb-cq}.

\section{Acknowledgement}

One of us (BB) thanks Prof. C.Quesne, PNTPM, University of Libre,
Brussels, for clarification of several points concerning the
infinite well problem, while another (AS) thanks Mr. Debayan Saha,
Indian Institute of Technology, Kharagpur for discussions and
helpful comments.


\pagebreak

\newpage


\noindent {\bf \large Figure Captions :}

\vspace{1cm}

\noindent {\bf Figure 1 :} \\
This shows a periodic trajectory for $n_y = 1 \ , \ n_x = 4$, so
that $ \tan \ \theta = \displaystyle \frac{p_y}{p_x} =
\frac{b}{4a} $.

\vspace{1cm}

\noindent {\bf Figure 2 :} \\
This shows a periodic trajectory for $n_y = 2 \ , \ n_x = 3$, so
that $ \tan \ \theta = \displaystyle \frac{p_y}{p_x} =
\frac{2b}{3a} $.

\vspace{1cm}

\noindent {\bf Figure 3 :} \\
This shows an open trajectory for $\theta = 30 ^0$.

\vspace{1cm}

\noindent {\bf Figure 4 :} \\
This shows some of the special trajectories for odd integral
values of both $n_y $ and $ n_x $, when the billiard ball hits one
of the corners and gets pocketed. The corners predicted in Table 1
agree with those actually traced out in this figure.

\vspace{1cm}

\noindent {\bf In each of the figures, the initial (starting)
trajectory is shown in blue, the intermediate ones in red, and the
final one in black.}

\end{document}